\documentclass[twocolumn,english,NOlinenumbers,aps,twocolumn,groupedaddress,superscriptaddress,showpacs]{revtex4-1}
\usepackage[T1]{fontenc}
\usepackage[latin9]{inputenc}
\usepackage{color}
\usepackage{babel}
\usepackage{amsmath}
\usepackage{graphicx}
\usepackage{esint}
\usepackage[unicode=true,pdfusetitle,
 bookmarks=true,bookmarksnumbered=false,bookmarksopen=false,
 breaklinks=false,pdfborder={0 0 0},backref=false,colorlinks=true]
 {hyperref}
\hypersetup{
 citecolor=blue}

\makeatletter

\usepackage{babel}

\makeatother

\begin{document}

\title{Dynamics of Localized Structures in Systems with Broken Parity Symmetry}

\author{J. Javaloyes}

\affiliation{Departament de Física, Universitat de les Illes Baleares, C/Valldemossa
km 7.5, 07122 Mallorca, Spain}

\author{P. Camelin}

\affiliation{Institut Non-Linéaire de Nice, Université de Nice Sophia Antipolis,
CNRS UMR 7335, 06560 Valbonne, France}

\author{M. Marconi}

\affiliation{Institut Non-Linéaire de Nice, Université de Nice Sophia Antipolis,
CNRS UMR 7335, 06560 Valbonne, France}

\author{M. Giudici}

\affiliation{Institut Non-Linéaire de Nice, Université de Nice Sophia Antipolis,
CNRS UMR 7335, 06560 Valbonne, France}
\begin{abstract}
A great variety of nonlinear dissipative systems are known to host
structures having a correlation range much shorter than the size of
the system. The dynamics of these Localized Structures (LSs) have
been investigated so far in situations featuring parity symmetry.
In this letter we extend this analysis to systems lacking of this
property. We show that the LS drifting speed in a parameter varying
landscape is not simply proportional to the parameter gradient, as
found in parity preserving situations. The symmetry breaking implies
a new contribution to the velocity field which is a function of the
parameter value, thus leading to a new paradigm for LSs manipulation.
We illustrate this general concept by studying the trajectories of
the LSs found in a passively mode-locked laser operated in the localization
regime. Moreover, the lack of parity affects significantly LSs interactions
which are governed by asymmetrical repulsive forces.

\end{abstract}
\maketitle
Symmetry breaking (SB) is undoubtedly one of the most important phenomenon
occurring in nature \cite{EB-PRL-64,TB-MOD-97,CGP-PRL-04}. One distinguishes
the situations of spontaneous SB, where the governing laws are symmetrical
but some of the solutions are not, from the cases of explicit SB where
the symmetry of the underlying theory is broken. In Optics, spontaneous
SB has been identified in pattern formation \cite{GMD-PRL-90}, coupled
nanocavities \cite{HHR-NAP-15}, ring lasers \cite{BGM-PRL-08} and
two dimensional \cite{SAL-PRA-14} or vectorial \cite{CSM-PRL-02}
solitons. Explicit SB is known to induce convective instabilities
\cite{CC-PRL-97,SCS-PRL-97,MN-PRL-00,WOT-PRE-00,MLA-PRL-08} and drifts
\cite{TBC-PRA-13,LMK-PRL-13,PGL-OL-14} and was recently studied in
PT symmetric waveguides \cite{BSS-PRA-12,ABS-PRA-12}. 

We address here dissipative extended systems hosting Localized Structures
(LSs) which are solutions characterized by a correlation length much
smaller than the size of the system making them individually addressable
\cite{TML-PRL-94,CRT-PRL-00,DC-LNP-11}. These states are ubiquitous
\cite{WKR-PRL-84,MFS-PRA-87,NAD-PSS-92,LMP-NAT-94,UMS-NAT-96,AP-PLA-01,1172836}
but they are particularly relevant for applications when implemented
in optical resonators and used as light bits for information processing
\cite{rosanov,FS-PRL-96,BLP-PRL-97,1172836}. Localized structures
were observed in the transverse \cite{BTB-NAT-02} and longitudinal
\cite{LCK-NAP-10,HBJ-NAP-14} dimensions of optical resonators driven
by an external field, in lasing cavities \cite{GBG-PRL-08,TAF-PRL-08}
and in optical parametric oscillators \cite{BWZ-PRE-07,BZV-PRE-11}.
Several theoretical paradigms have been used to describe these situations,
including the Lugiato-Lefever \cite{LL-PRL-87} and the Rosanov or
Ginzburg-Landau equations \cite{VFK-JOB-99}. All these systems share
invariance under the mirror symmetry of the variables in which localization
occurs. When LSs are implemented in the transverse section of semiconductor
microcavities, an additional equation accounts for carriers dynamics
but it still preserves the overall parity \cite{BTB-NAT-02} and the
hosted LSs are motionless when the underlying medium is homogeneous.
Because of translational invariance, LSs exhibit a Goldstone mode
\cite{FS-PRL-96,PhysRevE.66.046606} which is excited by any inhomogeneous
parameter variation, inducing their motion \cite{FGB-APL-06,PBC-APL-08,JEC-NAC-15}.
Since the velocity, instead of the acceleration, is proportional to
the parameter variations, the latter is interpreted as an Aristotelian
force. 

More recently, delayed systems have been analyzed from the perspective
of their equivalence with spatial extended systems \cite{GP-PRL-96}
and proposed for hosting LSs, fronts and chimera states \cite{GMZ-PRE-13,LPM-PRL-13,MGB-PRL-14,MJB-NAP-15,GJT-NC-15}.
In general, the non-instantaneous and causal response of the medium
implies a lack of parity in their spatiotemporal representation. In
this letter we show that, in an explicitly parity broken system, the
dynamics of temporal LSs in a modulated parameter landscape is fundamentally
different from the ones found in parity preserving situations. While
in the latter the motion of the LSs depends exclusively on the parameter
gradient, we reveal the existence in the former of another contribution
inducing a dependence of the velocity field on the local parameter
value. We consider an experimental situation where this contribution
is dominant and we formulate a new paradigm for LSs manipulation.
Moreover, we show that parity breaking leads to asymmetrical repelling
forces between LSs. 

In a parity broken system, traveling waves are the only possible solutions
and the behavior of LSs can be described considering a generic partial
differential equation (PDE) for a field $E\left(z,t\right)$
\begin{eqnarray}
\frac{\text{\ensuremath{\partial}}E}{\partial t}+\upsilon\left(\mu\right)\frac{\text{\ensuremath{\partial}}E}{\partial z} & = & \mu E+\mathcal{F}\left(|E|^{2},\frac{\partial^{2}}{\partial z^{2}}\right)E+Y.\label{eq:GenePDE}
\end{eqnarray}
with $\mathcal{F}$ a general nonlinear function whose explicit expression
is not needed. We note that equation (\ref{eq:GenePDE}) encompasses
the cases previously mentioned \cite{LL-PRL-87,VFK-JOB-99} and we
assume that it supports drifting LSs solutions defined as $E\left(z,t\right)=p\left(z-\upsilon t\right)$
with $p\left(u\right)$ an even function. For the sake of clarity,
Eq.(\ref{eq:GenePDE}) is written in a comoving frame such that $\upsilon\left(\mu_{0}\right)=0$.
Experimentally, when a single parameter is modulated, it may affect
several terms of the PDE and we consider the most general case where
the modulation affects an even and an odd derivative, i.e. the linear
gain $\mu$ and the drift velocity $\upsilon\left(\mu\right)$. The
influence of a small parameter variation $\mu=\mu_{0}+m\left(z\right)$
can be studied by a variational approach \cite{MBH-PRE-00} writing
$E=p\left[z-z_{0}\left(t\right)\right]$ which allows finding that
the position $z_{0}$ of the LS evolves according to 
\begin{eqnarray}
\frac{dz_{0}}{dt} & = & \left\{ \frac{\int p^{\dagger}\cdot pzdz}{-\int p^{\dagger}\cdot\dot{p}dz}\right\} \frac{dm}{dz}\left(z_{0}\right)+\left\{ \frac{d\upsilon}{d\mu}\left(\mu_{0}\right)\right\} m\left(z_{0}\right),\label{eq:Grad}
\end{eqnarray}
with $a\cdot b$ the dot product of complex numbers, and $p^{\dagger}$ and $\dot{p}$
two odd eigenfunctions representing the adjoint and direct neutral
translation modes. Because of the term $\left(d\upsilon/d\mu\right)$
coming from SB, the LS speed is proportional to the local value of
the parameter $m\left(z\right)$; if this term is neglected, one recovers
the case where LS motion is purely proportional to the gradient of
$m\left(z\right)$. 

These general considerations can be applied to the case of LSs hosted
in a laser field $E\left(z,t\right)$ coupled to a distant Saturable
Absorber (SA). While this scheme leads to conventional passive mode-locking
(PML) for cavity roundtrips $\tau$ shorter than the gain recovery
time $\tau<\tau_{g}$, we operate in the long cavity regime $\tau\gg\tau_{g}$,
and for bias currents below the lasing threshold of the compound system.
In this regime the PML pulses become \emph{localized}, i.e they become
lasing LSs \cite{MJB-PRL-14} which can be individually addressed
by an optical/electrical perturbation \cite{MJC-JSTQE-16}. By denoting
$G\left(z,t\right)$ the gain and $Q\left(z,t\right)$ the saturable
absorption, the Haus equations \cite{haus00rev} governing their dynamics
reads
\begin{eqnarray}
\negthickspace\negthickspace\frac{\partial E}{\partial t} & \negthickspace=\negthickspace & \left[\sqrt{\kappa}\left(1+\frac{1-i\alpha}{2}G-\frac{1-i\beta}{2}Q\right)-1+d\frac{\partial^{2}}{\partial z^{2}}\right]\negthickspace E,\label{eq:HF1}\\
\negthickspace\negthickspace\frac{\partial G}{\partial z} & \negthickspace=\negthickspace & \Gamma G_{0}-G\left(\Gamma+\left|E\right|^{2}\right),\frac{\partial Q}{\partial z}\negthickspace=\negthickspace Q_{0}-Q\left(1+s\left|E\right|^{2}\right),\label{eq:HF2}
\end{eqnarray}
where time has been normalized to the SA recovery time, $\alpha$
and $\beta$ are the linewidth enhancement factor of the gain and
absorber sections, $\kappa$ the fraction of the power remaining in
the cavity after each roundtrip, $G_{0}$ the pumping rate, $\Gamma=\tau_{g}^{-1}$
the gain recovery rate, $Q_{0}$ the modulation depth of the SA, $s$
the ratio of the saturation energy of the SA and of the gain sections
and $d=\left(2\gamma^{2}\right)^{-1}$ with $\gamma$ the bandwidth
of the spectral filter. In Eq.(\ref{eq:HF1}), the spatial variable
$z$ denotes the propagation along the cavity axis and corresponds
to a fast temporal scale for the LSs evolution within the cavity roundtrip
while $t$ is the slowly evolving time scale related to the LSs evolution
after each roundtrip. The steady states of Eqs.(\ref{eq:HF1},\ref{eq:HF2})
are the periodic solutions of the PML laser and Eqs.(\ref{eq:HF1},\ref{eq:HF2})
are subject to periodic boundary conditions $\left(E,G,Q\right)\left(z+\tau,t\right)=\left(E,G,Q\right)\left(z,t\right)$. 

\begin{figure}
\begin{centering}
\includegraphics[bb=20bp 0bp 450bp 300bp,clip,width=1\columnwidth]{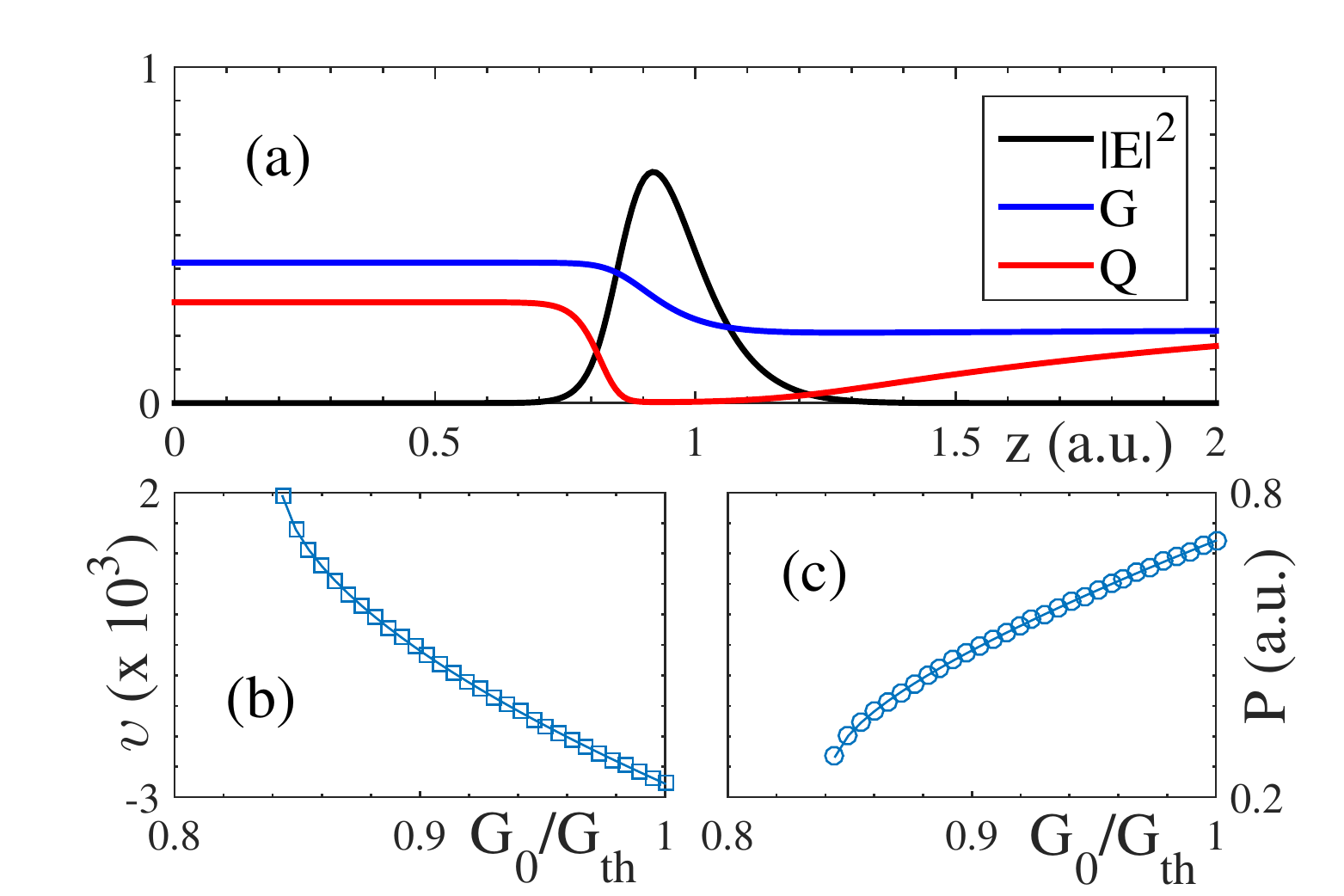}
\par\end{centering}

\centering{}\caption{(a) Profiles of the field intensity $\left|E\right|^{2}$, the carrier
$G$ and absorption $Q$. (b,c) Bifurcation diagram of the stable
solution branch for the drift velocity and the energy. Parameters
are $\left(\gamma,\kappa,\alpha,\beta,\Gamma,Q_{0},s\right)=\left(40,0.8,1,0.5,0.04,0.3,30\right)$.\label{Fig:theo1}}
\end{figure}

As the Eqs.(\ref{eq:HF2}) are only first order in $z$, LSs stem
here from a parity broken PDE and they exhibit a strong drift. The
adiabatic elimination of $G$ and $Q$ would cancel such a drift and
reduce Eq.(\ref{eq:HF1}) to the Ginzburg-Landau Equation. Although
in a semiconductor medium the adiabatic elimination of the absorption
may be sound, it cannot be applied to the gain because $\tau_{g}$
is much longer than the pulsewidth $\tau_{p}\sim\gamma^{-1}$. In
our case, the bias current $G_{0}$ controls at the same time the
linear gain, whose expression is $\mu=\sqrt{\kappa}(2+G_{0}-Q_{0})/2-1$
, but also higher order spatial derivatives due to the non-instantaneous
response of the medium. Assuming that $E\ll1$ and $\partial_{z}G_{0}\ll\Gamma G_{0}$,
$G\left(z\right)$ than can be expressed as 
\begin{eqnarray}
G\left(z\right) & \sim & G_{0}\left(z\right)\left[1-\int_{0}^{\infty}|E\left(z-r\right)|^{2}e^{-\Gamma r}dr\right].\label{eq:Gsol}
\end{eqnarray}

Expanding $E\left(z-r\right)$ in Taylor series in $r$ and integrating
over $r$ leads to an infinite series of even and odd derivatives
of $E$ with respect to $z$. The odd terms contribute to the drift
of the solutions and to SB. This establishes the conceptual link with
the generic situation depicted in Eq.\ref{eq:GenePDE}.

The Eqs.(\ref{eq:HF1}-\ref{eq:HF2}) were solved in \cite{MJB-PRL-14}
for the LSs energy; here we calculate also the LS drift velocity as
a function of the bias current. A typical asymmetrical profile can
be observed in Fig.\ref{Fig:theo1}(a) while the bifurcation diagrams
are depicted in Fig.\ref{Fig:theo1}(b,c). The solution branch for
the energy of the pulse $P=\int_{-\infty}^{\infty}\left|E\right|^{2}dz$
depicted in Fig.\ref{Fig:theo1}(c) shows the typical square root
behavior consistent with the fact that the LSs arise as a saddle-node
bifurcation of limit cycle \cite{MJB-PRL-14}. Importantly, one notices
that the drift velocity, or equivalently the deviation of the period
with respect to the roundtrip, is a strongly evolving function of
the bias current. The lack of the $\left(z\rightarrow-z\right)$ symmetry
is ultimately a consequence of the causality principle, because the
response of the medium is necessarily asymmetric with respect to an
intensity variation, as shown in Fig.\ref{Fig:theo1}(a) and Eq.\ref{eq:Gsol}.

Experimental evidence of this phenomenon is obtained using the setup
described in \cite{MJB-PRL-14,MJC-JSTQE-16} and by modulating the
pumping current of a VCSEL mounted in an external cavity closed by
a resonant saturable absorber mirror. When the modulation frequency
$\nu_{m}$ is almost resonant with the cavity free spectral range
$\nu_{c}$, i.e. small values of the detuning $\Delta=\nu_{m}-\nu_{c}$,
a quasi-stationary parameter variation is introduced inside the cavity.
In line with the theoretical analysis, LSs' dynamics in this parameter
landscape can be pictured using a pseudo spatio-temporal representation,
where the temporal trace is folded onto itself after a time that corresponds
to the cavity roundtrip $\nu_{c}^{-1}$. Accordingly, the roundtrip
number $n$ becomes a slow time variable proportional to $t$ while
the pseudo-space variable $z$ corresponds to the position within
the roundtrip \cite{GP-PRL-96}. However, to simplify the interpretation,
we present these diagrams in the reference frame of the modulation
signal, i.e. using a folding parameter $\nu_{m}^{-1}$.

\begin{figure}
\begin{centering}
\includegraphics[bb=200bp 0bp 3000bp 1142bp,clip,width=1\columnwidth]{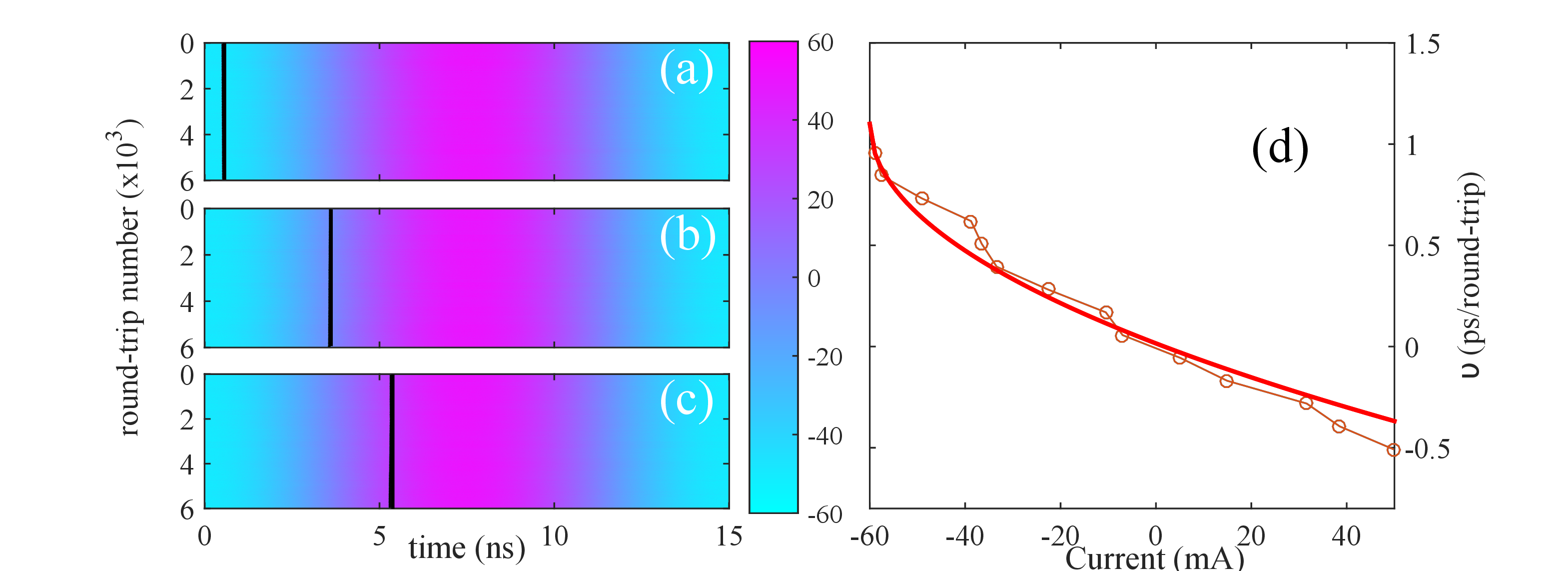}
\par\end{centering}

\caption{a-c) Spatio-temporal diagrams of the LS position evolution (black
line) when the current is sinusoidally modulated (color scale in mA)
with an amplitude $\delta J=120\,$mA around $J_{cw}=226\,$mA and
$\nu_{m}=66614250\,$Hz. The detunings are : a) $\Delta=-4.75\,$KHz,
b) $\Delta=-0.25\,$kHz and c) $\Delta=1.75\,$kHz. d) LS drifting
speed induced by the variation of $J$ (circle) and best fit with
a square-root function.  \label{vsJ}}
\end{figure}

We first modulate sinusoidally the pumping current around $J=J_{cw}$
and we represent in Fig.\ref{vsJ}(a-c) the evolution of LS position
on the current modulation landscape. When $\Delta=0$ the LS exhibits
a fixed position with respect to the modulation signal which is located
on the zero of the modulation on its positive slope. If $\left|\Delta\right|$
is small enough ($-4.75\,$kHz$<\Delta<2.5\,$kHz), the LS still sits
in a stationary position with respect the modulation signal. We shows
in Fig.\ref{vsJ}(a,c) that this position gets closer to the modulation
peak (resp. bottom) for increasing positive (resp. negative) value
of $\Delta$. It is worth noting that all stationary positions found
varying $\Delta$ are located on the positive slope of the modulation.
The presence of a finite $\Delta$ should induce a drifting speed
$\upsilon_{\Delta}$ which is given by $\upsilon_{\Delta}=\Delta\nu_{c}^{-2}$
in the limit of $\Delta\ll\nu_{c}$. The fact that the LS remains
locked to stationary positions in the modulation, as shown in Fig.\ref{vsJ}(a,c),
indicates that another Aristotelian force equilibrates $\upsilon_{\Delta}$.

These results can be explained by assuming that a current variation
around $J=J_{cw}$ induces a drifting speed $\upsilon_{J}$ of the
LS which depends on the value of $J$ instead of its time derivative,
as suggested by the theoretical analysis. At each stationary position
found for a given $\Delta\neq0$, $\upsilon_{\Delta}$ is compensated
by an opposite drifting speed induced by current variation $\upsilon_{J}$
such that $\upsilon_{\Delta}+\upsilon_{J}=0$. Accordingly, from the
value of the current at the stationary positions, it is possible to
establish the dependence of $\upsilon_{J}$ on $\left(J-J_{cw}\right)$,
as shown in Fig.\ref{vsJ}(d). In agreement with the theory, $\upsilon_{J}$
is a decreasing function of $J$. This explains why the stable equilibrium
points are located only on the positive slope of the modulation: J
plays a restoring (diverging) role on positive (negative) slope with
respect to deviations from the equilibrium point.

\begin{figure}
\centering{}\includegraphics[bb=45bp 0bp 707bp 186bp,clip,width=1\columnwidth]{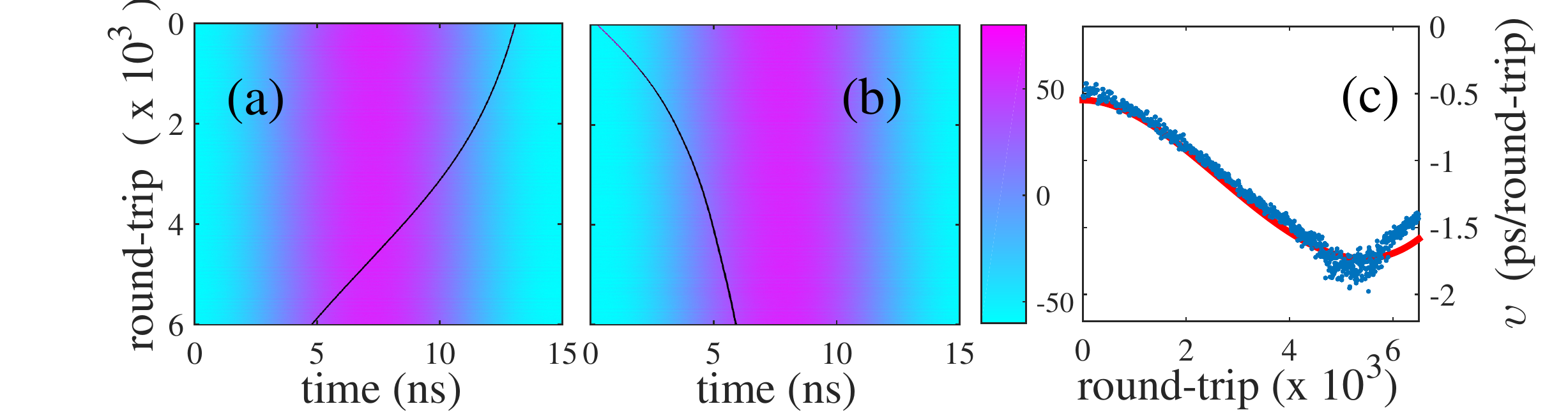}\caption{(a,b) Spatiotemporal diagrams of the LS position evolution (black
line) over the current modulation (color scale in mA). (a)$\Delta=-9.25\,$kHz,
(b) $\Delta=3.25\,$KHz, all other parameters as in Fig.~\ref{vsJ}.
(c) LS drifting speed (red) calculated from the derivative of the
trajectory shown in (a) and calculated as $\upsilon_{\Delta}+\upsilon_{J}$
(dots) with $\upsilon_{J}$ obtained fitting the curve $\upsilon_{J}$
in Fig.~\ref{vsJ}d) with $J$ the value of the current along the
trajectory. \label{dynamics}}
\end{figure}

For values of $\Delta$ outside the above specified interval, $\upsilon_{\Delta}$
cannot be balanced by $\upsilon_{J}$ at any current values spanned
by the modulation and the LS starts to drift in the space-time diagram,
as shown in Fig.\ref{dynamics}(a,b), in a way reminiscent of the
Adler unlocking mechanism of a forced oscillator. In the limit of
large detuning, the motion becomes uniform because $|\upsilon_{\Delta}|\gg|\upsilon_{J}|$.
From the LS time law, extracted from Fig.\ref{dynamics}(a), we infer
the instantaneous velocity of the LS in the space-time diagram, that
we represent in Fig.\ref{dynamics}(c), red curve. On the other hand,
the speed can also be calculated by adding $\upsilon_{\Delta}+\upsilon_{J}$,
where the first addendum is obtained from the value of $\Delta$ and
the second is obtained from the curve $\upsilon_{J}(J-J_{cw})$ plotted
in Fig.\ref{vsJ}(d), using the value of $J$ corresponding to the
position of the LS on the modulation landscape. The results obtained
are represented in Fig.\ref{dynamics}(c) using blue dots. The two
curves representing the instantaneous velocity coincide, thus indicating
that LS's speed depends only on the local current value, while the
derivative of the modulation signal is not playing any relevant role.

\begin{figure*}[t]
\begin{centering}
\includegraphics[bb=10bp 0bp 898bp 258bp,clip,width=2\columnwidth]{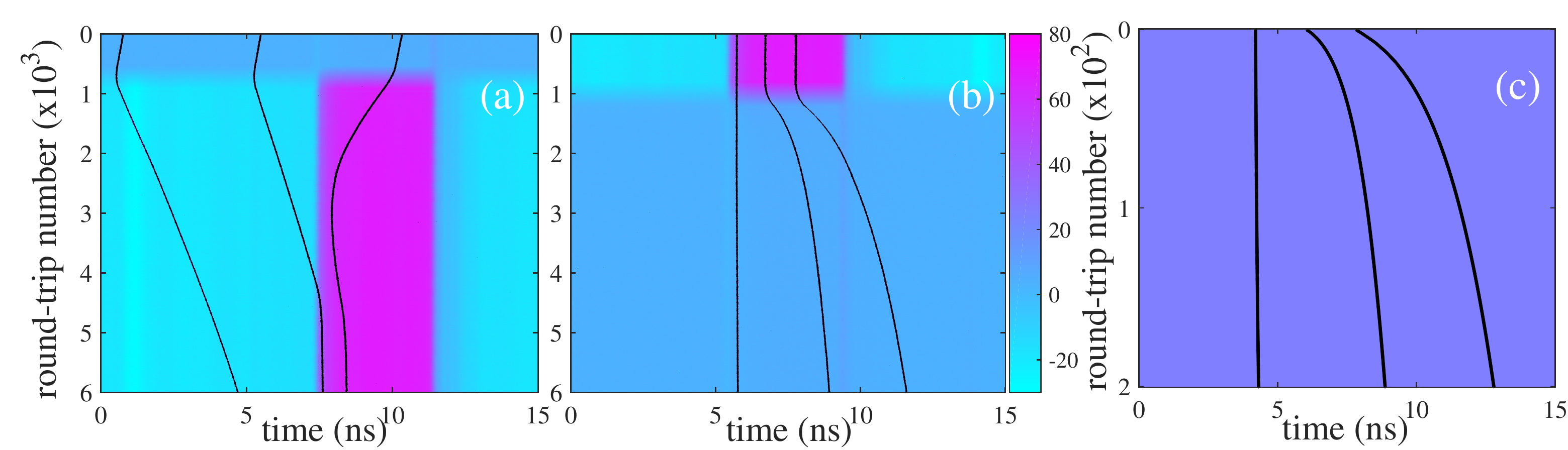} 
\par\end{centering}

\caption{Experimental spatiotemporal diagrams showing the evolution of three
LSs when an electrical square pulse of $4\,$ns is applied (a) and
removed (b) to the pumping current (color scale in mA). The current
pulse amplitude is $120\,$mA and $J_{cw}=220\,$mA. In (b) $\Delta=0$,
while $\Delta$ is slightly negative in (a) explaining the negative
drifting speed from $N=0$ until $N=800$. Panel c): Numerically calculated
trajectory using Eqs.~(\ref{eq:eqeff1},\ref{eq:eqeff2}), with parameters
$\left(\upsilon_{0},\Delta\upsilon,P_{0},\Delta P\right)=\left(0,-0.01,0.28,0.7\right)$,
$G_{sn}=0.845G_{th}$ and $\upsilon_{\Delta}=-2.35\times10^{-3}$.
Other parameters as in Fig.\ref{Fig:theo1}. \label{interaction}}
\end{figure*}

We realized another experiment using a bivalued rectangular current
modulation with upper and lower values $J_{u,l}$. Because a large
set of current values is spanned at the pulse rising edge, it becomes
an anchoring region for the LSs for a wide set of values of $\Delta$.
Yet, when $\Delta$ falls outside the locking interval, the LS drifts
in the space-time diagram and it exhibits two clearly different speeds
in the regions where $J=J_{u}$ and $J=J_{l}$. This gives further
evidence that LS speed does not depend on the time derivative of the
modulation, otherwise it would be identical on the two current plateaus.

When several LSs are present into the cavity, interaction forces come
into play and the current landscape enables their analysis. The LSs
evolution when the rectangular current modulation is suddenly applied
at roundtrip $N=800$ is shown in Fig.\ref{interaction}(a). Let us
identify the LSs from 1 to 3 using their position from left to right.
The LS$_{3}$ acquires a strong negative speed when passing on the
high current plateau as $J_{u}>J_{cw}$ while the two leftmost LSs
acquire a positive speed since $J_{l}<J_{cw}$. All the LSs try to
reach the rising edge of the modulation but, when LS$_{2}$ and LS$_{3}$
get too close at $N=4500$, repulsive interaction prevents LS$_{3}$
to occupy the stable equilibrium position where $\upsilon_{\Delta}+\upsilon_{J}=0$
, which is instead occupied by LS$_{2}$. Hence, LS$_{3}$ sits at
a position shifted of $0.8\,$ns, where the repulsive force generated
by LS$_{2}$ force balances the Aristotelian forces $\upsilon_{\Delta}+\upsilon_{J}(J_{u}-J_{cw})$.
This situation further evolves (not shown in Fig.\ref{interaction}(a))
when LS$_{1}$ reaches the rising front and pushes the other two LSs
to the right to occupy the position previously occupied by LS$_{2}$.
The three LSs then remain eventually gathered around the rising front
of the electrical pulse at a mutual distance of $0.8\,$ns. This situation
is similar to the one shown in Fig.\ref{interaction}(b) before the
modulation is removed at $N=1200$. Hereafter, the LSs evolve exclusively
under the action of repulsive forces and their dynamics reveal that
the action-reaction principle is violated as for instance LS$_{1}$
interacts with LS$_{2}$ but not vice-versa. The asymmetry of the
interaction is due to the broken parity and follows from the causality
principle.

Such a dynamics in a modulated parameter landscape can be theoretically
analyzed with the results of Fig.\ref{Fig:theo1}. Close to the saddle-node
bifurcation, we approximate the drift as a function of the current
as 
\begin{eqnarray}
\upsilon & =f_{\upsilon}\left(G_{0}\right)= & \upsilon_{0}+\Delta\upsilon\sqrt{G_{0}-G_{sn}},\label{eq:upsilon}
\end{eqnarray}
and similarly $P=f_{P}\left(G_{0}\right)$. The coefficients $\left(\upsilon_{0},\Delta\upsilon,P_{0},\Delta P\right)$
and $G_{sn}$ are given by the bifurcation diagram of a single LS
in Fig.\ref{Fig:theo1}(b,c). In the comoving frame of the external
periodic potential, the equation for the relative drift velocity Eq.(\ref{eq:upsilon})
transforms into $\tilde{\upsilon}+\upsilon_{\Delta}=f_{\upsilon}\left(G_{0}\right)$
with $\upsilon_{\Delta}$ the velocity of the drifting potential and
$\tilde{\upsilon}$ the residual speed. In this reference frame, the
external potential depends only on the fast time $z$ and not anymore
on the slow time $t$. As the characteristics of the PML pulses depend
only on the gain value at the leading edge $G^{\left(i\right)}$,
we replace in Eq.(\ref{eq:upsilon}) $G_{0}\rightarrow G^{\left(i\right)}$.
During the gain depletion occurring around the pulse, the so-called
fast stage, only the strongly nonlinear stimulated terms in Eqs.(\ref{eq:HF2})
are relevant and the gain at the falling edge of the LS is simply
$G^{\left(f\right)}=G^{\left(i\right)}\exp\left(-P\right)$. Provided
that the variations of the bias current $G_{0}\left(z\right)$ are
slower than $\tau_{g}$, the solution of the carrier equation in between
LSs reads
\begin{eqnarray}
G\left(z_{2}\right) & = & G\left(z_{1}\right)e^{-\Gamma D_{2}}+G_{0}\left(z_{2}\right)\left(1-e^{-\Gamma D_{2}}\right),
\end{eqnarray}
with $D{}_{n}=z_{n}-z_{n-1}$ and $G\left(z_{1}\right)$ an arbitrary
initial condition. By denoting as $z_{n}$ the position of the $n$-th
LS whose residual velocity is $\tilde{\upsilon}_{n}=dz_{n}/dt$ we
find that 
\begin{eqnarray}
\frac{dz_{n}}{dt} & = & f_{\upsilon}\left[G_{n}^{\left(i\right)}\right]-\upsilon_{\Delta}\;,\; P_{n}=f_{P}\left[G_{n}^{\left(i\right)}\right],\label{eq:eqeff1}\\
G_{n}^{\left(i\right)} & = & G_{n-1}^{\left(i\right)}e^{-P_{n-1}-\Gamma D_{n}}+G_{0}\left(z\right)\left(1-e^{-\Gamma D_{n}}\right).\label{eq:eqeff2}
\end{eqnarray}

For $N-$LSs with $n\in\left[1,\dots,N\right]$, the periodic boundary
conditions linking the gain depletion of the rightmost LS to the dynamics
of the leftmost is $z_{0}=z_{N}-\tau$. The repulsive interactions
mediated by the gain depletion is exemplified in Fig.\ref{interaction}c),
where, as shown in the experiment, LS$_{n}$ affects LS$_{n\text{+1}}$
but not vice-versa. The source of the asymmetry is visible in Eq.(\ref{eq:eqeff2})
and the interactions are repulsive since the velocity is a decreasing
function of the current $\left(\Delta\upsilon<0\right)$.

In conclusion, we described the dynamics of LSs in systems with an
explicitly broken parity symmetry appearing because of the causality
principle. Our analysis reveals that the Aristotelian forces ruling
their dynamical behavior are very different from the ones found in
parity preserving conditions. These results pave the way towards control
and manipulation of LSs for information processing in broken parity
systems and in particular to the control of three dimensional light
bullets in semiconductor lasers \cite{J-PRL-16}. 
\begin{acknowledgments}
J.J. acknowledges financial support from the Ramón y Cajal fellowship
and project COMBINA (TEC2015-65212-C3-3-P). The INLN Group acknowledges
funding of Région PACA with the Projet Volet Général 2011 GEDEPULSE
and ANR project OPTIROC. 
\end{acknowledgments}


\begin{thebibliography}{10}

\bibitem{EB-PRL-64}
F.~Englert and R.~Brout.
\newblock Broken symmetry and the mass of gauge vector mesons.
\newblock {\em Phys. Rev. Lett.}, 13:321--323, Aug 1964.

\bibitem{TB-MOD-97}
Patrick~P.L Tam and Richard~R Behringer.
\newblock Mouse gastrulation: the formation of a mammalian body plan.
\newblock {\em Mechanisms of Development}, 68(1-2):3 -- 25, 1997.

\bibitem{CGP-PRL-04}
Julyan H.~E. Cartwright, Juan~Manuel Garc\'{\i}a-Ruiz, Oreste Piro, C.~Ignacio
  Sainz-D\'{\i}az, and Idan Tuval.
\newblock Chiral symmetry breaking during crystallization: An
  advection-mediated nonlinear autocatalytic process.
\newblock {\em Phys. Rev. Lett.}, 93:035502, Jul 2004.

\bibitem{GMD-PRL-90}
C.~Green, G.~B. Mindlin, E.~J. D'Angelo, H.~G. Solari, and J.~R. Tredicce.
\newblock Spontaneous symmetry breaking in a laser: The experimental side.
\newblock {\em Phys. Rev. Lett.}, 65:3124--3127, Dec 1990.

\bibitem{HHR-NAP-15}
P.~Hamel, S.~Haddadi, F.~Raineri, P.~Monnier, G.~Beaudoin, I.~Sagnes,
  A.~Levenson, and A.~M. Yacomotti.
\newblock Spontaneous mirror-symmetry breaking in coupled photonic-crystal
  nanolasers.
\newblock {\em Nat Photon}, 9(5):311--315, May 2015.
\newblock Letter.

\bibitem{BGM-PRL-08}
S.~Beri, L.~Gelens, M.~Mestre, G.~Van der Sande, G.~Verschaffelt, A.~Scire,
  G.~Mezosi, M.~Sorel, and J.~Danckaert.
\newblock Topological insight into the non-arrhenius mode hopping of
  semiconductor ring lasers.
\newblock {\em Physical Review Letters}, 101(9):093903, 2008.

\bibitem{SAL-PRA-14}
V.~Skarka, N.~B. Aleksi\ifmmode~\acute{c}\else \'{c}\fi{},
  M.~Leki\ifmmode~\acute{c}\else \'{c}\fi{}, B.~N.
  Aleksi\ifmmode~\acute{c}\else \'{c}\fi{}, B.~A. Malomed, D.~Mihalache, and
  H.~Leblond.
\newblock Formation of complex two-dimensional dissipative solitons via
  spontaneous symmetry breaking.
\newblock {\em Phys. Rev. A}, 90:023845, Aug 2014.

\bibitem{CSM-PRL-02}
C.~Cambournac, T.~Sylvestre, H.~Maillotte, B.~Vanderlinden, P.~Kockaert, Ph.
  Emplit, and M.~Haelterman.
\newblock Symmetry-breaking instability of multimode vector solitons.
\newblock {\em Phys. Rev. Lett.}, 89:083901, Jul 2002.

\bibitem{CC-PRL-97}
A.~Couairon and J.~M. Chomaz.
\newblock Pattern selection in the presence of a cross flow.
\newblock {\em Phys. Rev. Lett.}, 79:2666--2669, Oct 1997.

\bibitem{SCS-PRL-97}
Marco Santagiustina, Pere Colet, Maxi San~Miguel, and Daniel Walgraef.
\newblock Noise-sustained convective structures in nonlinear optics.
\newblock {\em Phys. Rev. Lett.}, 79:3633--3636, Nov 1997.

\bibitem{MN-PRL-00}
Namiko Mitarai and Hiizu Nakanishi.
\newblock Spatiotemporal structure of traffic flow in a system with an open
  boundary.
\newblock {\em Phys. Rev. Lett.}, 85:1766--1769, Aug 2000.

\bibitem{WOT-PRE-00}
H.~Ward, M.~N. Ouarzazi, M.~Taki, and P.~Glorieux.
\newblock Influence of walkoff on pattern formation in nondegenerate optical
  parametric oscillators.
\newblock {\em Phys. Rev. E}, 63:016604, Dec 2000.

\bibitem{MLA-PRL-08}
A.~Mussot, E.~Louvergneaux, N.~Akhmediev, F.~Reynaud, L.~Delage, and M.~Taki.
\newblock Optical fiber systems are convectively unstable.
\newblock {\em Phys. Rev. Lett.}, 101:113904, Sep 2008.

\bibitem{TBC-PRA-13}
M.~Tlidi, L.~Bahloul, L.~Cherbi, A.~Hariz, and S.~Coulibaly.
\newblock Drift of dark cavity solitons in a photonic-crystal fiber resonator.
\newblock {\em Phys. Rev. A}, 88:035802, Sep 2013.

\bibitem{LMK-PRL-13}
Fran{\c{c}}ois Leo, Arnaud Mussot, Pascal Kockaert, Philippe Emplit, Marc
  Haelterman, and Majid Taki.
\newblock Nonlinear symmetry breaking induced by third-order dispersion in
  optical fiber cavities.
\newblock {\em Phys. Rev. Lett.}, 110:104103, Mar 2013.

\bibitem{PGL-OL-14}
Pedro Parra-Rivas, Dami\`{a} Gomila, Fran\c{c}ois Leo, St\'{e}phane Coen, and
  Lendert Gelens.
\newblock Third-order chromatic dispersion stabilizes kerr frequency combs.
\newblock {\em Opt. Lett.}, 39(10):2971--2974, May 2014.

\bibitem{BSS-PRA-12}
I.~V. Barashenkov, Sergey~V. Suchkov, Andrey~A. Sukhorukov, Sergey~V. Dmitriev,
  and Yuri~S. Kivshar.
\newblock Breathers in $\mathcal{PT}$-symmetric optical couplers.
\newblock {\em Phys. Rev. A}, 86:053809, Nov 2012.

\bibitem{ABS-PRA-12}
N.~V. Alexeeva, I.~V. Barashenkov, Andrey~A. Sukhorukov, and Yuri~S. Kivshar.
\newblock Optical solitons in $\mathcal{PT}$-symmetric nonlinear couplers with
  gain and loss.
\newblock {\em Phys. Rev. A}, 85:063837, Jun 2012.

\bibitem{TML-PRL-94}
M.~Tlidi, P.~Mandel, and R.~Lefever.
\newblock Localized structures and localized patterns in optical bistability.
\newblock {\em Phys. Rev. Lett.}, 73:640--643, Aug 1994.

\bibitem{CRT-PRL-00}
P.~Coullet, C.~Riera, and C.~Tresser.
\newblock Stable static localized structures in one dimension.
\newblock {\em Phys. Rev. Lett.}, 84:3069--3072, Apr 2000.

\bibitem{DC-LNP-11}
O.~Descalzi, M.~Clerc, S.~Residori, and G.~Assanto.
\newblock {\em Localized States in Physics: Solitons and Patterns}, volume 751
  of {\em Lecture Notes in Physics}.
\newblock Springer Berlin Heidelberg, 2011.

\bibitem{WKR-PRL-84}
J.~Wu, R.~Keolian, and I.~Rudnick.
\newblock Observation of a nonpropagating hydrodynamic soliton.
\newblock {\em Phys. Rev. Lett.}, 52:1421--1424, Apr 1984.

\bibitem{MFS-PRA-87}
E.~Moses, J.~Fineberg, and V.~Steinberg.
\newblock Multistability and confined traveling-wave patterns in a convecting
  binary mixture.
\newblock {\em Phys. Rev. A}, 35:2757--2760, Mar 1987.

\bibitem{NAD-PSS-92}
F.~J. Niedernostheide, M.~Arps, R.~Dohmen, H.~Willebrand, and H.~G. Purwins.
\newblock Spatial and spatio-temporal patterns in pnpn semiconductor devices.
\newblock {\em physica status solidi (b)}, 172(1):249--266, 1992.

\bibitem{LMP-NAT-94}
Kyoung-Jin Lee, William~D. McCormick, John Pearson, and Harry~L. Swinney.
\newblock Experimental observation of self-replicating spots in a
  reaction-diffusion system.
\newblock {\em Nature}, 369:215--218, 1994.

\bibitem{UMS-NAT-96}
P.~B. Umbanhowar, F.~Melo, and H.~L. Swinney.
\newblock Localized excitations in a vertically vibrated granular layer.
\newblock {\em Nature}, (382):793--796, 1996.

\bibitem{AP-PLA-01}
Yuri~A. Astrov and H.G. Purwins.
\newblock Plasma spots in a gas discharge system: birth, scattering and
  formation of molecules.
\newblock {\em Physics Letters A}, 283(5-6):349 -- 354, 2001.

\bibitem{1172836}
L.A. Lugiato.
\newblock Introduction to the feature section on cavity solitons: An overview.
\newblock {\em Quantum Electronics, IEEE Journal of}, 39(2):193--196, 2003.

\bibitem{rosanov}
N.~N. Rosanov and G.~V. Khodova.
\newblock Autosolitons in bistable interferometers.
\newblock {\em Opt. Spectrosc.}, 65:449, 1988.

\bibitem{FS-PRL-96}
W.~J. Firth and A.~J. Scroggie.
\newblock Optical bullet holes: Robust controllable localized states of a
  nonlinear cavity.
\newblock {\em Phys. Rev. Lett.}, 76:1623--1626, Mar 1996.

\bibitem{BLP-PRL-97}
M.~Brambilla, L.~A. Lugiato, F.~Prati, L.~Spinelli, and W.~J. Firth.
\newblock Spatial soliton pixels in semiconductor devices.
\newblock {\em Phys. Rev. Lett.}, 79:2042--2045, 1997.

\bibitem{BTB-NAT-02}
S.~Barland, J.~R. Tredicce, M.~Brambilla, L.~A. Lugiato, S.~Balle, M.~Giudici,
  T.~Maggipinto, L.~Spinelli, G.~Tissoni, T.~Kn{\"o}dl, M.~Miller, and
  R.~J{\"a}ger.
\newblock Cavity solitons as pixels in semiconductor microcavities.
\newblock {\em Nature}, 419(6908):699--702, Oct 2002.

\bibitem{LCK-NAP-10}
F.~Leo, S.~Coen, P.~Kockaert, S.P. Gorza, P.~Emplit, and M.~Haelterman.
\newblock Temporal cavity solitons in one-dimensional kerr media as bits in an
  all-optical buffer.
\newblock {\em Nat Photon}, 4(7):471--476, Jul 2010.

\bibitem{HBJ-NAP-14}
T.~Herr, V.~Brasch, J.~D. Jost, C.~Y. Wang, N.~M. Kondratiev, M.~L. Gorodetsky,
  and T.~J. Kippenberg.
\newblock Temporal solitons in optical microresonators.
\newblock {\em Nature Photonics}, 8(2):145--152, 2014.

\bibitem{GBG-PRL-08}
P.~Genevet, S.~Barland, M.~Giudici, and J.~R. Tredicce.
\newblock Cavity soliton laser based on mutually coupled semiconductor
  microresonators.
\newblock {\em Phys. Rev. Lett.}, 101:123905, Sep 2008.

\bibitem{TAF-PRL-08}
Y.~Tanguy, T.~Ackemann, W.~J. Firth, and R.~J\"ager.
\newblock Realization of a semiconductor-based cavity soliton laser.
\newblock {\em Phys. Rev. Lett.}, 100:013907, Jan 2008.

\bibitem{BWZ-PRE-07}
I.~V. Barashenkov, S.~R. Woodford, and E.~V. Zemlyanaya.
\newblock Interactions of parametrically driven dark solitons. i. n\'eel-n\'eel
  and bloch-bloch interactions.
\newblock {\em Phys. Rev. E}, 75:026604, Feb 2007.

\bibitem{BZV-PRE-11}
I.~V. Barashenkov, E.~V. Zemlyanaya, and T.~C. van Heerden.
\newblock Time-periodic solitons in a damped-driven nonlinear schr\"odinger
  equation.
\newblock {\em Phys. Rev. E}, 83:056609, May 2011.

\bibitem{LL-PRL-87}
L.~A. Lugiato and R.~Lefever.
\newblock Spatial dissipative structures in passive optical systems.
\newblock {\em Phys. Rev. Lett.}, 58:2209--2211, May 1987.

\bibitem{VFK-JOB-99}
A.~G. Vladimirov, S.~V. Fedorov, N.~A. Kaliteevskii, G.~V. Khodova, and N.~N.
  Rosanov.
\newblock Numerical investigation of laser localized structures.
\newblock {\em Journal of Optics B: Quantum and Semiclassical Optics},
  1(1):101, 1999.

\bibitem{PhysRevE.66.046606}
J.~M. McSloy, W.~J. Firth, G.~K. Harkness, and G.-L. Oppo.
\newblock Computationally determined existence and stability of transverse
  structures. ii. multipeaked cavity solitons.
\newblock {\em Phys. Rev. E}, 66:046606, Oct 2002.

\bibitem{FGB-APL-06}
F.~Pedaci, P.~Genevet, S.~Barland, M.~Giudici, and J.~R. Tredicce.
\newblock Positioning cavity solitons with a phase mask.
\newblock {\em Applied Physics Letters}, 89(22):221111, 2006.

\bibitem{PBC-APL-08}
F.~Pedaci, S.~Barland, E.~Caboche, P.~Genevet, M.~Giudici, J.~R. Tredicce,
  T.~Ackemann, A.~J. Scroggie, W.~J. Firth, G.-L. Oppo, G.~Tissoni, and
  R.~J\"{a}ger.
\newblock All-optical delay line using semiconductor cavity solitons.
\newblock {\em Applied Physics Letters}, 92(1):011101, 2008.

\bibitem{JEC-NAC-15}
Jae~K. Jang, Miro Erkintalo, Stephane Coen, and Stuart~G. Murdoch.
\newblock Temporal tweezing of light through the trapping and manipulation of
  temporal cavity solitons.
\newblock {\em Nat Commun}, 6, Jun 2015.
\newblock Article.

\bibitem{GP-PRL-96}
G.~Giacomelli and A.~Politi.
\newblock Relationship between delayed and spatially extended dynamical
  systems.
\newblock {\em Phys. Rev. Lett.}, 76:2686--2689, Apr 1996.

\bibitem{GMZ-PRE-13}
Giovanni Giacomelli, Francesco Marino, Michael~A. Zaks, and Serhiy Yanchuk.
\newblock Nucleation in bistable dynamical systems with long delay.
\newblock {\em Phys. Rev. E}, 88:062920, Dec 2013.

\bibitem{LPM-PRL-13}
Laurent Larger, Bogdan Penkovsky, and Yuri Maistrenko.
\newblock Virtual chimera states for delayed-feedback systems.
\newblock {\em Phys. Rev. Lett.}, 111:054103, Aug 2013.

\bibitem{MGB-PRL-14}
Francesco Marino, Giovanni Giacomelli, and Stephane Barland.
\newblock Front pinning and localized states analogues in long-delayed bistable
  systems.
\newblock {\em Phys. Rev. Lett.}, 112:103901, Mar 2014.

\bibitem{MJB-NAP-15}
M.~Marconi, J.~Javaloyes, S.~Balle, and M.~Giudici.
\newblock Vectorial dissipative solitons in vertical-cavity surface-emitting
  lasers with delays.
\newblock {\em Nature Photonics}, 2015.

\bibitem{GJT-NC-15}
B.~Garbin, J.~Javaloyes, G.~Tissoni, and S.~Barland.
\newblock Topological solitons as addressable phase bits in a driven laser.
\newblock {\em Nat. Com.}, 6, 2015.

\bibitem{MBH-PRE-00}
T.~Maggipinto, M.~Brambilla, G.~K. Harkness, and W.~J. Firth.
\newblock Cavity solitons in semiconductor microresonators: Existence,
  stability, and dynamical properties.
\newblock {\em Phys. Rev. E}, 62:8726--8739, Dec 2000.

\bibitem{MJB-PRL-14}
M.~Marconi, J.~Javaloyes, S.~Balle, and M.~Giudici.
\newblock How lasing localized structures evolve out of passive mode locking.
\newblock {\em Phys. Rev. Lett.}, 112:223901, Jun 2014.

\bibitem{MJC-JSTQE-16}
M.~Marconi, J.~Javaloyes, P.~Camelin, D.~Chaparro, S.~Balle, and M.~Giudici.
\newblock Control and generation of localized pulses in passively mode-locked
  semiconductor lasers.
\newblock {\em Selected Topics in Quantum Electronics, IEEE Journal of},
  PP(99):1--1, 2015.

\bibitem{haus00rev}
H.~A. Haus.
\newblock Mode-locking of lasers.
\newblock {\em IEEE J. Selected Topics Quantum Electron.}, 6:1173--1185, 2000.

\bibitem{J-PRL-16}
J.~Javaloyes.
\newblock Cavity light bullets in passively mode-locked semiconductor lasers.
\newblock {\em Phys. Rev. Lett.}, 116:043901, Jan 2016.

\end{thebibliography}

\end{document}